# Imaging of Spin Dynamics in Closure Domain and Vortex Structures

J. P. Park, P. Eames, D. M. Engebretson, J. Berezovsky, and P. A. Crowell

*School of Physics and Astronomy, University of Minnesota, 116 Church St. SE*
*Minneapolis, MN 55455*

Abstract

Time-resolved Kerr microscopy is used to study the excitations of individual micron-scale ferromagnetic thin film elements in their remnant state. Thin (18 nm) square elements with edge dimensions between 1 and 10 µm form closure domain structures with 90 degree Néel walls between domains. We identify two classes of excitations in these systems. The first corresponds to precession of the magnetization about the local demagnetizing field in each quadrant, while the second excitation is localized in the domain walls. Two modes are also identified in ferromagnetic disks with thicknesses of 60 nm and diameters from 2 µm down to 500 nm. The equilibrium state of each disk is a vortex with a singularity at the center. As in the squares, the higher frequency mode is due to precession about the internal field, but in this case the lower frequency mode corresponds to gyrotropic motion of the entire vortex. These results demonstrate clearly the existence of well-defined excitations in inhomogeneously magnetized microstructures.

PACS numbers: 75.40.Gb, 75.75.+a, 75.60.-d

The excitation spectra of ferromagnetic films have been probed in detail using tools such as ferromagnetic resonance,[1] and Brillouin light scattering.[2]   In the past several years,  there has been an increasing focus on  patterned ferromagnetic films with transverse dimensions on the order of 10 microns or less.[2-6]  In this limit, there is not necessarily a clear distinction between exchange-dominated excitations (spin-waves) and magnetostatic modes.   Even less is known about the appropriate description of excitations in mesoscopic systems that are not homogeneously magnetized.    In systems with negligible magnetocrystalline anisotropy,  the balance of demagnetization and exchange energies results in flux closure structures such as the four-fold closure domain pattern in thin squares[7] or the vortex state of disks.[8, 9]  These types of simple domain structures are ubiquitous in the size regime just above the single-domain limit, and a detailed knowledge of their excitation spectra will be essential in achieving a complete understanding of nanoscale magnetic structures.

We have studied the excitation spectra of individual ferromagnetic thin film structures with in-plane dimensions down to 500 nm using time-resolved Kerr microscopy as a local spectroscopic probe.   We identify several unique modes of these systems, including excitations localized in domain walls and the gyrotropic mode of a single vortex in submicron diameter disks.[10]   The combination of spatial and time resolution  allows us to distinguish between simple precessional modes and the more interesting excitations that arise directly from geometric confinement or the presence of domain walls. Micromagnetic simulations show strong qualitative agreement with the experiments, which have been carried out for permalloy ($Ni_{0.81}Fe_{0.19}$)  squares with edge dimensions between 1 and 10 μm  and disks with diameters  between 500 nm and 2 μm.

The thin film structures for the measurements discussed here were prepared by electron beam lithography and lift-off of $Ni_{0.81}Fe_{0.19}$ films sputtered on GaAs (100) substrates. The thickness of the square samples was 18 nm, and structures with edge dimensions of 10, 5, 3, 2, and 1 micron were studied. Circles with thicknesses of 60 nm and diameters of 2, 1, and 0.5 µm were also prepared. After polishing of the GaAs substrate to a thickness of approximately 25 µm, each sample was placed on a 30 µm wide section of a tapered stripline and positioned under a 100X oil immersion objective with a numerical aperture of 1.25. The time-resolved Kerr microscopy technique used for this work has been described previously by several groups.[11-13] A 76 MHz pulse train of 150 femtosecond pulses from a Ti:sapphire laser is split into pump and probe beams. The pump pulse is incident on a fast photodiode which converts the optical pulse into a current pulse that is discharged into the stripline. The resulting magnetic field pulse at the sample is in the plane of the film [see Fig. 1(a)] and has an amplitude of approximately 5 Oe and a temporal width of approximately 150 psec. The polar Kerr rotation of the time-delayed probe beam is measured using a polarization bridge. The output of the bridge is read by a lock-in amplifier referenced to a chopper in the pump beam path, so that the measured signal is the change in the z-component of the magnetization due to the pump pulse. The objective is scanned using a piezoelectric stage, and a polar Kerr image of the sample is collected at each time delay. All measurements discussed in this paper were carried out in zero field after reducing the field from saturation.

A polar Kerr image of a 5 µm permalloy square during the pump pulse is shown in the inset of Fig. 1(b). Images at later times are shown in the insets of Figs. 1(c) and (d). As for all of the square samples discussed in this paper, the remnant state is a simple closure domain structure, with four quadrants separated by 90 degree Néel walls. The pulsed field is oriented vertically in the plane of the image, and so the spins in the top and bottom quadrants experience torques of opposite sign while the two side quadrants experience no torque during the pulse. As a result, the observed $M_z$ during the pulse in Fig. 1(b) is positive in the top quadrant, negative in the bottom quadrant, and zero elsewhere. The left panels of Figs. 1 (b), (c), and (d) show the time evolution of the polar Kerr rotation at the three positions on the sample indicated by the dots shown in each inset. The respective Fourier transforms are shown in the right-hand panels. The time traces from the top and bottom quadrants differ only by a 180 degree phase shift, while the scan taken near a domain wall shows a distinctly lower oscillation frequency. The two frequencies, 1.8 and 0.8 GHz, observed in Figs. 1(b), (c), and (d) represent the two dominant modes of the system. This is confirmed in Fig. 1(e), which shows the average of the power spectra obtained at all positions on the square.

The data of Fig. 1 suggest that the lower frequency mode is associated with the domain walls. To examine this question more closely, we have constructed frequency-domain images of the polar Kerr signal by calculating the Fourier transform of the time-domain data at each position. As expected from Fig. 1(a), the spectral weight is concentrated near 0.8 and 1.8 GHz, and images at these two frequencies are shown in Fig. 2(a). The higher frequency mode is clearly concentrated in the center of the top and bottom domains, although some spectral power appears in the two side quadrants. This mode has nodes along the domain walls and near the center-line of the sample. The node at the center-line is required by symmetry, since the response of the top and bottom halves of the sample must be 180 degrees out of phase. In contrast to the higher frequency mode, the spectral power at 0.8 GHz is concentrated in the domain walls between the quadrants. A similar two-mode structure is observed for two, three, and ten micron squares. (The lifetime observed in the one micron square was too short to resolve the mode structure.)

Some insight into the nature of the two modes is gained through micromagnetic simulations, which we have performed by integration of the Landau-Lifshitz-Gilbert equation

$$(1+\alpha^2)\frac{\partial \mathbf{M}_i}{\partial t} = -\gamma(\mathbf{M}_i \times \mathbf{H}_{eff,i}) - \frac{\gamma\alpha}{M_S}\mathbf{M}_i \times (\mathbf{M}_i \times \mathbf{H}_{eff,i}), \quad (1)$$

where $\mathbf{H}_{eff}$ is the total effective field, which in this case includes the demagnetizing, exchange, and pulsed fields, and $\mathbf{M}_i$ is the magnetization in each cell. We have used the parameters $\gamma/2\pi$ = 2.95 GHz/kOe, exchange constant $A = 1.3 \times 10^{-6}$ erg/cm, magnetization $M_s$ = 700 emu/cm$^3$,[13] and thickness $d$ = 17.5 nm. The calculations were performed using the Object-Oriented Micromagnetic Framework (OOMMF).[14] Each sample was discretized into 400 × 400 × 1 cells and relaxed into its initial state with a damping constant $\alpha$ = 0.5. The damping constant was then set to a more realistic value of 0.008 and the pulse was applied. The simulated time-domain data were convolved with an optical resolution function represented by a Gaussian with a FWHM of 540 nm (corresponding to the measured resolution) and resampled on a grid matched to the experimental pixel size. A Fourier transform was then applied to the output in order to produce frequency-domain images. Simulated spectral images at the dominant frequencies are shown for a 5 μm square in Fig. 2 (b). As in the experiment, two modes are observed in the simulations, with power concentrated in the top and bottom domains and domain walls, respectively. The frequencies of the two modes as a function of size are shown along with the experimental values in Fig. 2(c). Although the frequency scale in the simulations is higher, both the spectral images and size dependence are in good qualitative agreement with the experimental results.

Some insight into the physical difference between the two types of modes can be obtained by examining the effective field acting on the spins in the square. The micromagnetic effective field, including the demagnetization and exchange fields, is shown in Fig. 3 as a function of azimuthal angle at a radius of 1.25 μm from the center of the 5 μm square. The demagnetization field makes the dominant contribution to the effective field, even in the domain wall, as is evident from the much smaller scale for the exchange field shown in Fig. 3. The total effective field varies slowly in the center of each domain, and the dynamic response is essentially uniform precession about the local demagnetizing field. This mode has almost purely magnetostatic character and is not fundamentally different from the response of a uniformly magnetized specimen to a spatially inhomogeneous microwave field.[15] The domain wall modes, however, exist in a region of rapidly varying effective field, and it is less clear how to deduce the observed resonant frequencies from a simple physical argument. One approach is motivated by previous work on domain wall dynamics. Argyle *et al.*[16] studied similar closure domain structures in garnet films at larger length scales and lower frequencies, applying a model in which the vortex that exists at the intersection of the four Néel walls is subject to a restoring force originating from the magnetic poles formed when the walls are displaced. The resonant frequency we calculate based on their model is about a factor of eight smaller than the observed value for the 5 μm square. Another possibility is that the inhomogeneous demagnetizing field creates an effective potential well for spin-waves as found recently for ferromagnetic wires,[17] [13] but the lower symmetry in the current problem makes an analytical treatment of the spin-wave localization problem difficult.

Finally, we consider a somewhat simpler relative of the closure domain structure: a magnetic vortex. In cylindrical nanoparticles with thicknesses of the order of the exchange length $L_E = \sqrt{A/M_S^2}$ (~ 18 nm for permalloy) and aspect ratios $\beta = L/R \sim 0.1 - 0.5$, where $L$ is the thickness and $R$ the radius, the magnetic ground state is a vortex in which the magnetization curls around the central axis.[8] [18] The core of the vortex is formed by the central singularity at which the magnetization points out of the plane of the film. A schematic of the magnetization in this structure is shown in Fig. 4(a) along with a magnetic force microscope (MFM) image of a cylindrical permalloy disk with a diameter of 500 nm and a thickness of 60 nm. We have studied the dynamical response of disks with diameters of 2 µm, 1 µm, and 500 nm after each was subjected to a 150 psec field pulse. Each of these disks formed a single vortex in zero field as determined by MFM measurements.

The time-domain polar Kerr signal obtained at positions near the center of each disk is shown for the three different diameters in Figs. 4 (b), (c), and (d) along with results from the corresponding simulations, which were carried out on 400 × 400 × 1 grids in which the circles were inscribed. The parameters for the simulations were identical to those defined above for the squares. The higher frequency signal that appears in each case is attributed to precession of the magnetization about the local internal field. (Although a high-frequency signal from the 500 nm disk is not evident in Fig. 4(d), it can be observed in the Fourier transform.) However, the extremely long-lived low-frequency signal distinguishes these data from the response observed in the case of the closure domain structures. The frequency of this mode increases with decreasing diameter.  Although the frequency (0.6 ± 0.1 GHz) of the low-frequency mode for the 500 nm disk is of the same order as for the domain wall modes seen in squares, the observed lifetime is significantly longer, as can be seen by comparing the time traces of Fig. 4 with those of Fig. 1(c).

Excitations of vortices in sub-micron disks have been investigated theoretically by Guslienko and co-workers,[10] starting from an equation of motion derived by Thiele.[19] A moving vortex experiences a Magnus force perpendicular to velocity, and it therefore undergoes a spiral motion as it approaches the equilibrium position after an in-plane magnetic field pulse.  Guslienko *et al.* show that the frequency of this gyrotropic motion is

$$\omega_0 = \frac{1}{2}\gamma M_S \frac{\xi^2}{\chi(0)}, \qquad (2)$$

where γ is the gyromagnetic ratio and $M_S$ is the saturation magnetization. The parameter $\xi$ and the susceptibility $\chi(0)$ depend on the magnetization distribution in the displaced vortex.  The simplest model assumes that the entire vortex moves rigidly, although this requires free poles to exist at the edges of the disk.  It was shown in Ref. [10] that an alternative model that avoids edge poles[18] gives eigenfrequencies calculated from Eq. 2 that are significantly closer to those determined from a full micromagnetic calculation.  In our case, both the experimental and simulated frequencies are closer to the "pole-free" model, within 20 % for the 1 μm and 500 nm disks, than the rigid vortex result, for which the discrepancy is at least 50 %.[10]  We note that although the gyrotropic motion of a vortex has been inferred previously from the analysis of domain wall resonance in closure domain structures,[16] the data of Fig. 4 represent the first direct observation of this mode in an isolated vortex.

In summary, we have identified the low-lying excitations of inhomogeneously magnetized microstructures, including the simple closure domain structure in squares and the vortex state of disks down to 500 nm in diameter. A domain-wall mode in the squares and a gyrotropic mode of the vortices are identified in addition to more conventional precessional modes at higher frequencies. Although the qualitative agreement between experiment and micromagnetic simulations is good, there remain a number of outstanding questions, including the appropriate physical description of the domain-wall modes as well as the origin of the observed damping times.

This work was supported by NSF DMR 99-83777, the Research Corporation, the Alfred P. Sloan Foundation, the University of Minnesota MRSEC (DMR 98-09364), and the Minnesota Supercomputing Institute. We acknowledge helpful discussions with C. E. Campbell and M. Yan.

**Figure Captions**

Fig. 1: (a) Schematic of the experiment, showing the orientation of the pulsed magnetic field. All measurements were made in zero static field after the samples were demagnetized. (b), (c), (d) The polar Kerr signal measured as a function of pump-probe delay (left) and its Fourier transform (right) for a 5 µm permalloy square at the three locations on the sample indicated in the insets. Each inset shows a polar Kerr image of the sample at the times indicated, with 0 psec corresponding to the peak of the pump pulse. White and black indicate positive and negative signal respectively. (e) The average of the frequency power spectrum over the entire sample, showing peaks at 0.8 and 1.8 GHz.

Fig. 2: (a) Experimental spectral images of the response of a 5 µm square at the two frequencies corresponding to the peaks in the average spectrum of Fig. 1(e). (b) Spectral images obtained from the Landau-Lifshitz-Gilbert simulation using the parameters described in the text. The frequencies shown correspond to the peaks in the simulated spectra, which are slightly higher than their experimental counterparts. (c) Values of the domain center (closed symbols) and domain wall (open symbols) frequencies as a function of square size. Experimental and simulation results are shown by squares and circles respectively.

Fig. 3: Total static effective field (solid curve), including demagnetization and exchange fields, in the remnant state of the 5 µm square, shown for points around the perimeter of a circle at a radius of 1.25 µm from the center of the square. Only the circumferential

component of the field is shown. The sharp negative peaks correspond to domain walls. The exchange contribution to the effective field is shown as the dotted curve.

Fig. 4: (a) Schematic of a vortex structure (left) and a magnetic force microscope image (right) of a 500 nm disk. The bright spot at the center of the disk in the image is due to the large z-component of the magnetization. (b), (c), (d) Experimental (left) and simulated (right) time-domain polar Kerr signals for vortex structures of diameters 2 μm, 1 μm, and 500 nm near the center of each disk. The low-frequency signal that is particularly prominent in the case of the 500 nm disk is the gyrotropic mode discussed in the text.

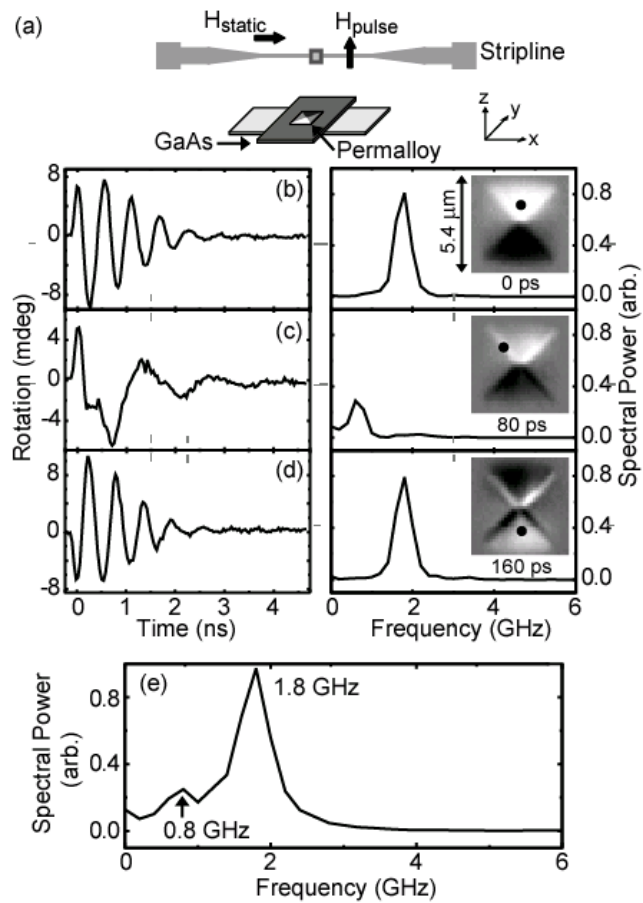

Fig. 1: Park *et al.*

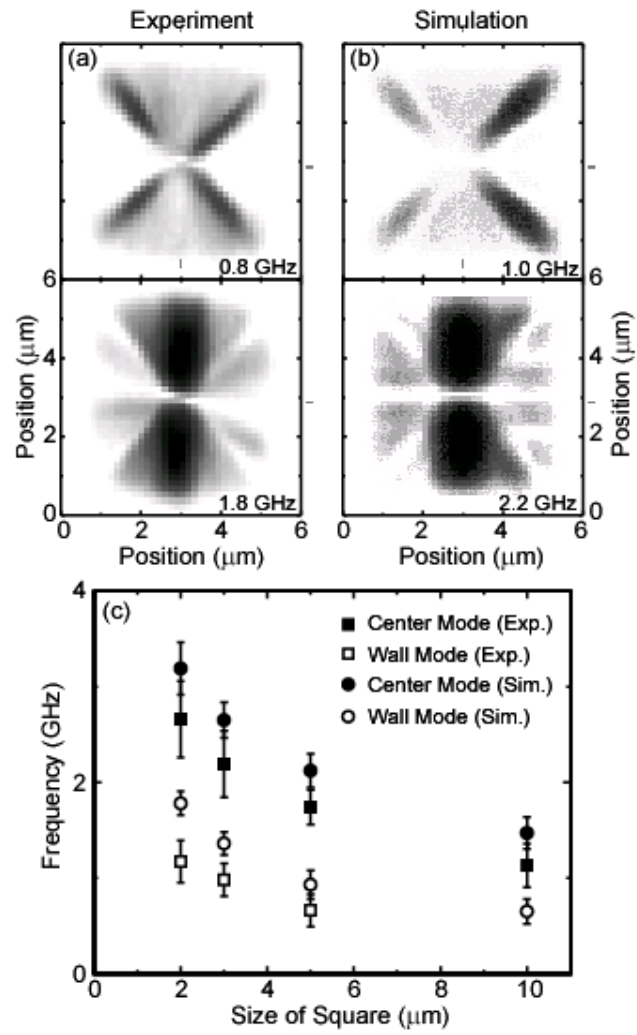

Fig. 2: Park *et al.*

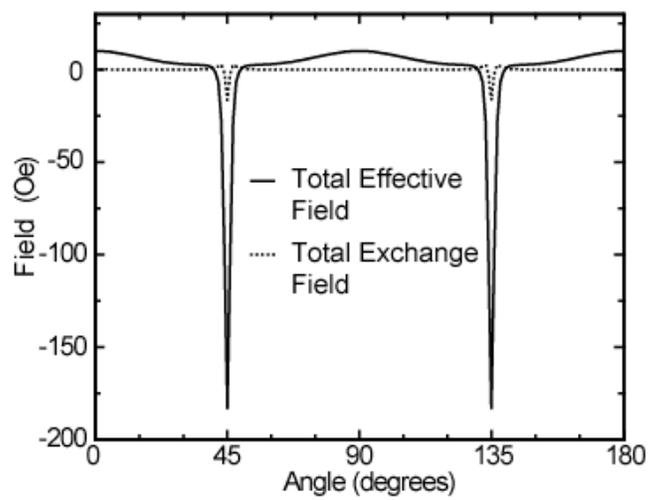

Fig. 3: Park *et al.*

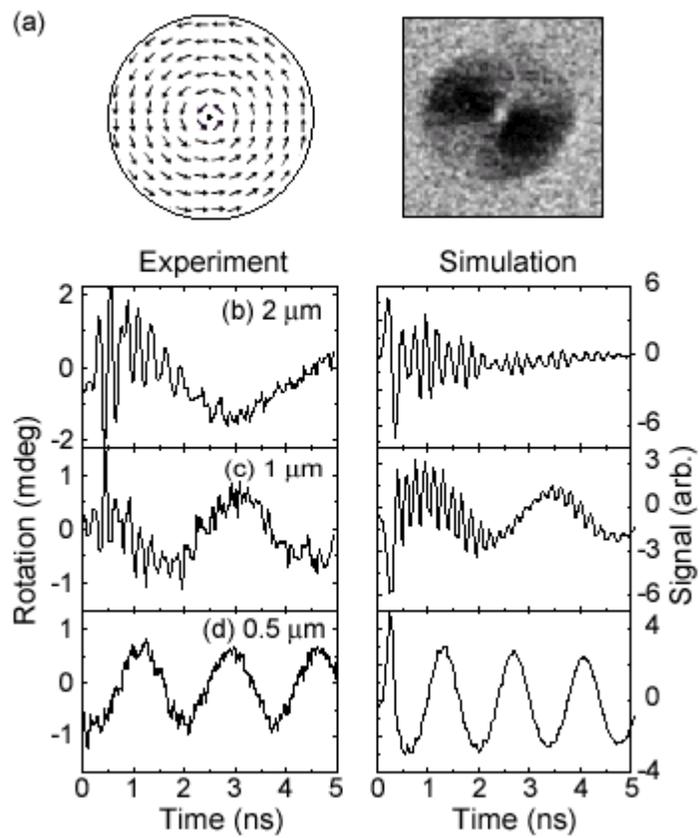

Fig. 4: Park *et al.*